\documentclass[runningheads,a4paper]{llncs}
\pdfoutput=1
\usepackage{amssymb}
\usepackage{graphicx}
\usepackage{amsmath}
\usepackage{url}

\begin{document}

\title{Search Personalization with Embeddings\thanks{{\scriptsize In Proceedings of the 39th European Conference on Information Retrieval, ECIR 2017, to appear.}}}
\titlerunning{Search Personalization with Embeddings}

%
%
\author{Thanh Vu\inst{1} \and Dat Quoc Nguyen\inst{2} \and Mark Johnson\inst{2} \and Dawei Song\inst{1} \and Alistair Willis\inst{1}}

\institute{The Open University, Milton Keynes, United Kingdom\\
\email{\{thanh.vu,dawei.song,alistair.willis\}@open.ac.uk}
\and
Department of Computing, Macquarie University, Sydney, Australia \\
\email{dat.nguyen@students.mq.edu.au,mark.johnson@mq.edu.au}
}

\authorrunning{{Thanh Vu, Dat Quoc Nguyen, Mark Johnson, Dawei Song,  Alistair Willis}}

%
%
\maketitle

\begin{abstract}
\vspace{-10pt}
Recent research has shown that the performance of search personalization depends on the richness of user profiles which normally represent the user's topical interests. In this paper, we propose a new embedding approach to learning user profiles, where users are embedded on a topical interest space. We then directly utilize the user profiles for search personalization. Experiments on query logs from a major commercial web search engine demonstrate that our embedding approach improves the performance of the search engine and also achieves better search performance than other strong baselines.
\vspace{-10pt}
\end{abstract}

\section{Introduction}
Users' personal data, such as a user's historic interaction with the search engine (e.g., submitted queries, clicked documents), have been shown useful to personalize search results to the  users' information need \cite{Bennett2012,WhiteE2013}. Crucial to effective search personalization is the construction of user profiles to represent individual users' interests \cite{Bennett2012,Cheng2016,HarveyB2013,Liu2015,Vu2014}. A common approach is to use  main topics discussed in the user's clicked documents \cite{Bennett2012,HarveyB2013,Vu2014,WhiteE2013}, which can be obtained by using a human generated ontology as in \cite{Bennett2012,WhiteE2013} or using an unsupervised topic modeling technique as in \cite{HarveyB2013,Vu2014}.

However, using the user profile to directly personalize a search has been not very successful with a \emph{minor} improvement \cite{HarveyB2013,Vu2014} or even \emph{deteriorate} the search performance \cite{HarveyB2011}. The reason is that each user profile is normally built using only the user's relevant documents (e.g., clicked documents),  ignoring user interest-dependent information related to input queries.   Alternatively, the user profile is utilized as a feature of a multi-feature learning-to-rank (L2R) framework \cite{Bennett2012,Vu2017,Vu2015,WhiteE2013}. In this case, apart from the user profile, dozens of other features has been proposed as the input of an L2R algorithm \cite{Bennett2012}. Despite being successful in improving  search quality, the contribution of the user profile is not very clear. 

To handle these problems, in this paper, we propose a new \emph{embedding} approach to constructing a user profile, using both the user's  input queries and relevant documents. We represent each user profile  using two projection matrices and a user embedding. The two projection matrices is to identify the user interest-dependent aspects of input queries and relevant documents while the user embedding is to capture the relationship between the queries and documents in this user interest-dependent subspace. 
 We then \emph{directly} utilize the user profile to re-rank the search results returned by a commercial search engine. Experiments on the query logs of a commercial web search engine demonstrate that modeling user profile with embeddings helps to significantly improve the performance of the search engine and also achieve better results than other comparative baselines \cite{Bennett2012,Teevan2011,Vu2015} do.

\section{Our approach}

We start with our new embedding approach to building user profiles in Section \ref{ssec:profile}, using pre-learned document embeddings and query embeddings. We then detail the processes of using an unsupervised topic model (i.e., Latent Dirichlet Allocation (LDA) \cite{Blei2003}) to learn document embeddings and query embeddings in Sections \ref{ssec:topics} and \ref{ssec:query}, respectively. We finally use the user profiles to personalize the search results returned by a commercial search engine in Section \ref{ssec:rank}.

\subsection{Building user profiles with embeddings}
\label{ssec:profile}

Let $\mathcal{Q}$ denote the set of queries, $\mathcal{U}$ be the set of users, and $\mathcal{D}$ be the set of documents. Let $(q, u, d)$ represent a triple $(\mathsf{query, user, document})$. The query $q \in \mathcal{Q}$, user $u \in \mathcal{U}$ and document $d \in \mathcal{D}$ are represented by vector embeddings $\boldsymbol{v}_q$, $\boldsymbol{v}_u$ and $\boldsymbol{v}_d \in\mathbb{R}^{k}$, respectively. 
 
Our goal is to select a \textit{score function} $f$ such that the implausibility value $f(q, u, d)$ of a correct triple $(q, u, d)$ (i.e. $d$ is a relevant document of $u$ given $q$) is \textit{smaller} than the implausibility value $f(q', u', d')$ of an incorrect triple $(q', u', d')$ (i.e. $d'$ is not a relevant document of $u'$ given $q'$). Inspired by embedding models of entities and relationships in knowledge bases \cite{NguyenCoNLL2016,Nguyen2016NAACL},  the score function $f$ is defined as follows:

\vspace{-10pt}
\begin{equation}
f(q, u, d) = \| \textbf{W}_{u,1}\boldsymbol{v}_q + \boldsymbol{v}_u - \textbf{W}_{u,2}\boldsymbol{v}_d \|_{\ell_{1/2}} 
\label{equa:stranse}
\end{equation}

\noindent here we represent the profile for the user $u$ by two matrices $\textbf{W}_{u,1}$ and $\textbf{W}_{u,2} \in \mathbb{R}^{k\times k}$ and a vector embedding $\boldsymbol{v}_u$, which represents the user's topical interests. Specifically, we use the interest-specific matrices $\textbf{W}_{u,1}$ and $\textbf{W}_{u,2}$ to identify the interest-dependent aspects of both query $q$ and document $d$, and use vector   $\boldsymbol{v}_u$ to describe the relationship between $q$ and $d$ in this interest-dependent subspace. 

In this paper, $\boldsymbol{v}_d$ and $\boldsymbol{v}_q$ are pre-determined by employing the LDA topic model \cite{Blei2003}, which are detailed in next Sections \ref{ssec:topics} and \ref{ssec:query}. Our model parameters are only the user embeddings $\boldsymbol{v}_u$ and matrices $\textbf{W}_{u,1}$ and $\textbf{W}_{u,2}$. To learn these user embeddings and matrices, we minimize the margin-based objective function:

\vspace{-5pt}
\begin{equation}
\mathcal{L} = \sum_{\substack{(q,u,d) \in \mathcal{G} \\ (q',u,d') \in \mathcal{G}'_{(q,u,d)}}} \max\big(0, \gamma + f(q, u, d) - f(q', u, d')\big)
\end{equation}

\noindent where $\gamma$ is the margin hyper-parameter, $\mathcal{G}$ is the training set that contains only correct triples, and $\mathcal{G}'_{(q, u, d)}$ is the set of incorrect triples generated by corrupting the correct triple $(q,u,d)$ (i.e. replacing the relevant document/query $d/q$ in $(q,u,d)$ by irrelevant documents/queries $d'/q'$). 
We use Stochastic Gradient Descent (SGD) to minimize $\mathcal{L}$, and
impose the following
constraints during training:
$\|\boldsymbol{v}_u\|_2 \leqslant 1$, $\| \textbf{W}_{u,1}\boldsymbol{v}_q\|_2
\leqslant 1$ and $\| \textbf{W}_{u,2}\boldsymbol{v}_d \|_2 \leqslant 1$. First, we initialize user matrices as identity matrices and then fix them to only learn the randomly initialized user embeddings. Then in the next step, we fine-tune the user embeddings and user matrices together. In all experiments shown in Section \ref{sec:expsetup}, we train for 200 epochs during each two
optimization step.

\subsection{Using LDA to learn document embeddings}
\label{ssec:topics}

In this paper, we model document embeddings by using topics extracted from relevant documents. 
We use LDA \cite{Blei2003} to \emph{automatically} learn $k$ topics from the  relevant document collection. After training an LDA model to calculate the probability distribution over topics for each document, we use the topic proportion vector of each document as its document embedding. Specifically, the $z^{th}$ element ($z = 1,2,...,k$) of the vector embedding for document $d$ is: $\boldsymbol{v}_{d,z} = \mathrm{P}(z \mid d)$ where $\mathrm{P}(z \mid d)$ is the probability of the topic $z$ given the document $d$.

\subsection{Modeling search queries with embeddings}
\label{ssec:query}

We also represent each query as a probability distribution $\boldsymbol{v}_{q}$ over topics, i.e. the $z^{th}$ element of the vector embedding for query $q$ is  defined as: $\boldsymbol{v}_{q,z} = \mathrm{P}(z \mid q)$ where $\mathrm{P}(z \mid q)$ is the probability of the topic $z$ given the query $q$. 
 Following \cite{Bennett2012,Vu2015}, we define $\mathrm{P}(z \mid q)$  as a mixture of LDA topic probabilities of $z$ given documents related to $q$. Let $\mathcal{D}_q = \{d_1, d_2, ..., d_n\}$ be the set of top $n$ ranked documents returned for a query $q$ (in the experiments we select $n=10$). We define $\mathrm{P}(z \mid q)$ as follows:

\vspace{-5pt}
\begin{equation}
\mathrm{P}(z \mid q) = \sum\nolimits_{i=1}^{n} \lambda_i \mathrm{P}(z \mid d_i)
\end{equation} 

\noindent where $\lambda_i = \frac{\delta^{{i} - 1}}{\sum_{j=1}^{n}\delta^{{j} - 1}}$ is the exponential decay function of $i$ which is the rank of $d_i$ in $D_q$. And $\delta$ is the decay hyper-parameter ($0 < \delta < 1$). The decay function is to specify the fact that a higher ranked document is more relevant to user in term of the lexical matching (i.e. we set the larger mixture weights to higher ranked documents).

\subsection{Personalizing search results}
\label{ssec:rank}

We utilize the user profiles (i.e., the learned user embeddings and matrices) to re-rank the original list of documents produced by a commercial search engine as follows:  (1) We download the top $n$ ranked documents given the input query $q$. We denote a downloaded document as $d$.  (2) For each document $d$ we apply the trained LDA model to infer the topic distribution $\boldsymbol{v}_{d}$. We then model the query $q$ as a topic distribution $\boldsymbol{v}_{q}$ as in Section \ref{ssec:query}.  (3) For each triple $(q, u, d)$, we calculate the implausibility value $f(q, u, d)$ as defined in Equation \ref{equa:stranse}. We then sort the values in the ascending order to achieve a new ranked list.

\section{Experimental methodology}
\label{sec:expsetup}

{\textbf{Dataset:}} We evaluate our new approach using the search results returned by a commercial search engine. We use a dataset of  query logs of 
of 106 anonymous users in 15 days from 01 July 2012 to 15 July 2012. 
A log entity contains a user identifier, a query, top-$10$ URLs ranked by the search engine, and clicked URLs along with the user's dwell time. We also download the content documents of these URLs for training LDA \cite{Blei2003} to learn document and query embeddings (Sections \ref{ssec:topics} and \ref{ssec:query}).

Bennett \emph{et al.} \cite{Bennett2012} indicate that short-term (i.e. session) profiles achieved better search performance than the longer-term profiles. Short-term profiles are usually constructed using the user's search interactions within a search session and used to personalize the search within the session \cite{Bennett2012}. To identify a search session, we use 30 minutes of user inactivity to demarcate the session boundary. In our experiments, we build short-term profiles and utilize the profiles to personalize the returned results. Specifically, we uniformly separate the last log entries within search sessions into a \textit{test set} and a \textit{validation set}. The remainder of log entities within search sessions are used for \textit{training} (e.g. to learn user embeddings and matrices in our approach). 

 \noindent {\textbf{Evaluation methodology:}} We use the SAT criteria detailed in \cite{FoxE2005} to identify whether a clicked URL is relevant from the query logs (i.e., a SAT click). That is either a click with a dwell time of at least 30 seconds or the last result click in a search {session}. We assign a positive (relevant) label to a returned URL if it is a SAT click. The remainder of the top-10 URLs is assigned negative (irrelevant) labels. We use the rank positions of the positive labeled URLs as the ground truth to evaluate the search performance before and after re-ranking. We also apply a simple pre-processing on these datasets as follows. At first, we remove the queries whose positive label set is empty from the dataset. After that, we discard the domain-related queries (e.g. Facebook, Youtube). To this end, the training set consists of 5,658 correct triples. The test and validation sets contain 1,210 and 1,184 correct triples, respectively. Table \ref{table:1} presents the dataset statistics after pre-processing.

\vspace{-5pt}
\begin{table}[ht]
\centering
\caption{Basic statistics of the dataset after pre-processing}
\label{table:1}
\begin{tabular}{c|c|c|c|c|c} \hline
\#days &\#users & \#distinct queries & \#SAT clicks & \#sessions &\#distinct documents \\ \hline
15 & 106 & 6,632 & 8,052 & 2,394 & 33,591\\
\hline
\end{tabular}
\end{table}
 \vspace{-5pt}

 \noindent{\textbf{Evaluation metrics}}: We use two standard evaluation metrics in document ranking \cite{Bennett2012,Manning2008}: \emph{mean reciprocal rank} (\textbf{MRR}) and \emph{precision} (\textbf{P@1}). 
For each metric, the higher value indicates the better ranking performance.

 \noindent {\textbf{Baselines:}} We employ three comparative baselines with the same experimental setup: (1) \textbf{SE}: The original rank from the search engine (2) \textbf{CI}: We promote returned documents previously clicked by the user. This baseline is similar to the personalized navigation method in Teevan \emph{et al.} \cite{Teevan2011}. (3) \textbf{SP}: The search personalization method using the short-term profile \cite{Bennett2012,Vu2015}. These are very comparative baselines given that they start with the ranking provided by the major search engine and add other signals (e.g., clicked documents) to get a better ranking performance \cite{Teevan2011,Bennett2012}. 

 \noindent {\textbf{Hyper-parameter tuning:}} 
We perform a grid search to select optimal hyper-parameters on the validation set. 
 We train the LDA model\footnote{We use the LDA implementation in Mallet toolkit: \url{http://mallet.cs.umass.edu/}.} using only the relevant documents (i.e., SAT clicks) extracted from the query logs, with the number of topics (i.e. the number of vector dimensions) $k \in \{50, 100, 200\}$. We then apply the trained LDA model to infer document embeddings and query embeddings for all documents and queries. We then choose either the $\ell_1$ or $\ell_2$ norm in the score function $f$, and select SGD learning rate $\eta \in \{0.001, 0.005, 0.01\}$, the margin hyper-parameter $\gamma \in \{1, 3, 5\}$ and the decay hyper-parameter $\delta \in \{0.7, 0.8, 0.9\}$. The highest MRR on the validation set is obtained when using $k = 200$, $\ell_1$ in $f$, $\eta = 0.005$, $\gamma = 5$, and $\delta = 0.8$.

\section{Experimental results}

Table \ref{tb1} shows the performances of the baselines and our proposed method. Using the previously clicked documents \emph{CI} helps to significantly improve the search performance ($p < 0.05$ with the \emph{paired t-test}) with the relative improvements of about 7$^+$\% in both MRR and P@1 metrics. With the use of short-term profiles as a feature of a learning-to-rank framework, \emph{SP} \cite{Bennett2012,Vu2015} improves the MRR score over the original rank significantly ($p < 0.01$) and achieves a better performance than \emph{CI}'s. 

\vspace{-10pt}
\begin{table}[ht]
\centering
\caption{Overall performances of the methods in the test set. \textbf{Our method}$_{-W}$ denotes the simplified version of our method. The subscripts denote the relative improvement over the baseline \emph{SE}.}
\label{tb1}
\begin{tabular}{l|c|c|c|c|c} \hline
Metric & \textbf{SE} & \textbf{CI} \cite{Teevan2011} & \textbf{SP} \cite{Bennett2012,Vu2015} & \textbf{Our method} & \textbf{Our method}$_{-W}$\\
\hline
{MRR} & 0.559 & 0.597$_{+6.9\%}$ & 0.631$_{+12.9\%}$ & \textbf{0.656$_{+17.3\%}$} & 0.645$_{+15.4\%}$\\\hline
{P@1} & 0.385 & 0.416$_{+8.1\%}$ & 0.452$_{+17.4\%}$ & \textbf{0.501$_{+30.3\%}$} & 0.481$_{+24.9\%}$\\ 
\hline
\end{tabular}
\end{table}
\vspace{-10pt}

By directly learning user profiles and applying them to re-rank the search results, our embedding approach achieves the highest performance of search personalization. Specifically, our MRR score is significantly ($p < 0.05$) higher than that of \emph{SP} (with the relative improvement of 4\% over SP). Likewise, the P@1 score obtained by our approach is significantly higher than that of the baseline \emph{SP} ($p < 0.01$) with the relative improvement of 11\%. 

In Table \ref{tb1}, we also present the performances of a simplified version of our embedding approach where we fix the user matrices as identity matrices and then only learn the user embeddings. Table \ref{tb1} shows that our simplified version achieves second highest scores compared to all others.\footnote{Our approach obtains significantly higher P@1 score ($p < 0.05$) than our simplified version with 4\% relative improvement.} Specifically, our simplified version obtains significantly higher P@1 score (with $p < 0.05$) than \emph{SP}.

\section{Conclusions}
In this paper, we propose a new embedding approach to building user profiles. We model each user profile using a user embedding together with two user matrices. The user embedding and matrices are then learned using LDA-based vector embeddings of the user's relevant documents and submitted queries. Applying it to web search, we use the profile to re-rank search results returned by a commercial web search engine. Our experimental results show that the proposed method can stably and significantly improve the ranking quality. 

\medskip
\noindent \textbf{Acknowledgments}:  The first two authors contributed equally to this work. Dat Quoc Nguyen is supported by an International Postgraduate Research Scholarship and a NICTA NRPA Top-Up Scholarship. 

\bibliographystyle{abbrv}
{
\bibliography{sigproc}}
\end{document}